\documentclass[twocolumn]{article}
\usepackage{times,bm}
\usepackage{amsmath}
\usepackage[affil-it,]{authblk}
\usepackage[pagebackref=false,colorlinks=true,linkcolor=acsblue,citecolor=olive,urlcolor=acsblue]{hyperref}

\usepackage{geometry}

\usepackage[switch*, displaymath,pagewise, mathlines]{lineno}
\leftlinenumbers
\usepackage{graphicx}
\usepackage{cite}
\usepackage{amsmath}
\usepackage{amsfonts}
\usepackage{amssymb}
\usepackage{slashed}
\usepackage{subfloat}
\usepackage{slashed}
\usepackage{color}
\usepackage{float}
\usepackage{multirow,multicol}
\usepackage{tabulary}
\usepackage[flushleft]{threeparttable}
\usepackage{pgfplots,subfigure}
\usepackage{xcolor}
\numberwithin{equation}{section}
\usepackage{tikz}
\usepackage{ulem}
\usepackage{hyperref}
\usepackage{nameref}
\usepackage{comment}

\usepgflibrary{arrows}
\usetikzlibrary{shapes.callouts}
\tikzset{
	level/.style   = { thick, },
	connect/.style = { dotted, red   },
	notice/.style  = { draw, rectangle callout, callout relative pointer={#1} },
	label/.style   = { text width=1cm }
}

\definecolor{acsblue}{RGB}{17,76,139}
\definecolor{shadecolor}{RGB}{255,241,204}
\usepackage{letltxmacro}
\makeatletter
\let\oldr@@t\r@@t
\def\r@@t#1#2{%
	\setbox0=\hbox{$\oldr@@t#1{#2\,}$}\dimen0=\ht0
	\advance\dimen0-0.2\ht0
	\setbox2=\hbox{\vrule height\ht0 depth -\dimen0}%
	{\box0\lower0.4pt\box2}}
\LetLtxMacro{\oldsqrt}{\sqrt}
\renewcommand*{\sqrt}[2][\ ]{\oldsqrt[#1]{#2}}
\makeatother
\usepackage{environ}

\begin{document}

\newcommand{{\ri}}{{\rm{i}}}
\newcommand{{\Psibar}}{{\bar{\Psi}}}
\newcommand*\var{\mathit}
%
%
\fontsize{8}{9}\selectfont
\title{\mdseries{Coupled fermion-antifermion pairs within a traversable wormhole}}
\author{ \textit {\mdseries{ Abdullah Guvendi}}$^{\ 1}$\footnote{\textit{ E-mail: abdullah.guvendi@erzurum.edu.tr \href{https://orcid.org/0000-0003-0564-9899}{(0000-0003-0564-9899)}}}~,~ \textit {\mdseries{ Omar Mustafa}}$^{\ 2}$\footnote{\textit{ E-mail: omar.mustafa@emu.edu.tr \href{https://orcid.org/0000-0001-6664-3859}{(0000-0001-6664-3859)}} }~,~ \textit {\mdseries{ Semra Gurtas Dogan}}$^{\ 3}$\footnote{\textit{ E-mail: semragurtasdogan@hakkari.edu.tr \textcolor{olive}{(Corr. Author)}\\
\href{https://orcid.org/0000-0001-7345-3287}{(0000-0001-7345-3287)}} } \\
	\small \textit {$^{\ 1}$ \footnotesize Department of Basic Sciences, Erzurum Technical University, 25050, Erzurum, Turkiye}\\
	\small \textit {$^{\ 2}$ \footnotesize Department of Physics, Eastern Mediterranean University, 99628, G. Magusa, north Cyprus, Mersin 10 - Turkiye}\\
	\small \textit {$^{\ 3}$ \footnotesize Department of Medical Imaging Techniques, Hakkari University, 30000, Hakkari, Turkiye}}

\date{}

\maketitle

\begin{abstract}
This study investigates the dynamics of fermion-antifermion ($f\overline{f}$) pairs within a traversable wormhole (TWH) spacetime by solving the two-body covariant Dirac equation with a position-dependent mass \(m \to m(r)\). In the context of a static, radially symmetric (2+1)-dimensional TWH characterized by a constant redshift function and a given shape function, we explore two Lorentz scalar potentials: (i) a Coulomb-like potential and (ii) an exponentially decaying potential. The Coulomb potential leads to positronium-like binding energies, with the ground state (\(n=0\)) energy approximately \(\mathcal{E}^{b}_{n} \approx -m_{e}c^2\alpha^2/4 \sim -6.803\) eV. On the other hand, the exponential potential establishes critical mass thresholds, \(m_c = \frac{(n + \frac{1}{2})\hbar}{2 \lambda_c c}\), at which the energy approaches zero, causing the system to cease to exist over time. Stability is maintained when \(n + \frac{1}{2} < 2\), resulting in oscillatory behavior, while \(n + \frac{1}{2} > 2\) leads to decay. The energy spectrum reveals essential features of the system, and the wave function reflects the influence of the wormhole’s throat, shaping spatial configurations and probability distributions. This work enhances our understanding of quantum phenomena in curved spacetimes and establishes connections to condensed matter physics.
\end{abstract}

\begin{small}
\begin{center}
\end{center}
\end{small}

\section{\mdseries{Introduction}}\label{sec:1}

In recent decades, wormholes have garnered considerable interest as potential solutions to Einstein's field equations. The profound impact of gravitational fields on quantum mechanical systems implies that wormholes, also referred to as Einstein-Rosen (ER) bridges, could hold significant cosmological importance \cite{R1.1}. Current advancements in physics do not preclude the possibility that advanced civilizations might harness wormholes for interstellar travel or even time travel \cite{R1, R5,R5.1}. Wormholes in (2+1)-dimensional spacetime offer not only educational value but also a valuable framework for exploring the quantization of gravitational fields \cite{R5.2}. This framework facilitates an in-depth investigation into several fascinating properties of wormhole spacetimes, including the Cauchy horizon \cite{R5.1}, wormhole stability \cite{R5.3}, and tests of wormhole solutions in extended gravity theories \cite{R5.31, R5.32, R5.33}, as well as frameworks like Rastall-Rainbow gravity \cite{R5.34, R5.35, R5.36} and 4D Einstein-Gauss-Bonnet (EGB) gravity \cite{R5.37}. In the realm of quantum gravity, efforts to unify quantum mechanics with general relativity have prompted detailed studies of quantum particles and antiparticles within wormhole spacetimes. These investigations reveal that the intense gravitational fields surrounding the wormholes create substantial spacetime curvature, significantly affecting the dynamics of quantum particles \cite{R5.4, R5.5, R5.6, R5.7, R5.8, R5.9, R5.10, R5.11}. Such effects may manifest as phenomena like time dilation and the alteration of particle trajectories \cite{R5.3}, among other intriguing behaviors.

\vspace{0.2cm}
\setlength{\parindent}{0pt}

Unlike black holes, which are characterized by non-removable singularities, wormholes offer a pathway connecting disparate regions of the universe. These connections may link two separate universes—commonly referred to as the upper and lower universes \cite{R5}—or join distant regions within the same universe, all while avoiding the presence of a singularity \cite{R1}. The geometry of a wormhole is defined by a non-zero minimum radial distance \( r_{\min} = \ell_{\circ} \), which corresponds to the throat radius. Nevertheless, the wormhole spacetime metric retains a singularity at the radial coordinate \( r_{\min} = \ell_{\circ} \). For a (2+1)-dimensional TWH spacetime to satisfy the Morris-Thorne criteria for traversability, the metric must be both radially symmetric and static \cite{R5, R6}. Within this framework, and assuming natural units where \( c = 1 = \hbar = G \), the metric of a (2+1)-TWH is given as follows \cite{R1, R5, R6, R7}:  
\begin{equation}
ds^2 = -e^{2\Phi(r)} dt^2 + \frac{dr^2}{1 - \frac{b(r)}{r}} + r^2 d\phi^2,\label{eq1:I}
\end{equation}  
where \( \Phi(r) \) and \( b(r) \) denote the redshift function and the form/shape function, respectively, both of which are arbitrary radial functions. In this study, we specifically analyze the case where \( \Phi'(r) = 0 \), implying that the redshift function is constant. This assumption significantly simplifies the field equations, enabling the derivation of novel solutions distinct from those predicted by general relativity. We adopt the shape function form \( b(r) = \frac{\ell_{\circ}^2}{r} \), which characterizes a negative energy density in general relativity. In this scenario, the TWH metric simplifies to the following form:  
\begin{equation}
ds^2 = -dt^2 + d\rho^2 + \mathcal{R}^2(\rho) d\phi^2, \quad \mathcal{R}(\rho) = \sqrt{\rho^2 + \ell_{\circ}^2},\label{eq2:I}
\end{equation}  
where \( \rho^2 = r^2 - \ell_{\circ}^2 \), and \( \ell_{\circ} \) represents the wormhole throat radius (as discussed in \cite{wh1, wh2}). This metric describes a minimal surface, called a catenoid, which serves as a material analog for a wormhole, specifically the catenoid bridge \cite{AS}. The catenoid, as a minimal surface, resembles two-dimensional wormholes and has been proposed as a potential model for TWHs \cite{wh2} (see also \cite{gibbons}). Its curvature approaches zero at infinity, while the curvature is concentrated around the bridge region \cite{n-5}. The catenoid structure has numerous potential applications, such as serving as a bridge between graphene layers \cite{wh2, n-6} or nanotubes. The impact of the catenoid surface on fermionic dynamics has been explored in both non-relativistic \cite{n-6} and relativistic contexts \cite{yesiltas}. Much of this research has concentrated on single-particle test fields, though many-body systems remain a challenging area. Nevertheless, investigating $f\overline{f}$ pairs within this spacetime could yield significant insights. Non-perturbative solutions for these coupled systems in curved spacetime may provide valuable information about their optical and electronic characteristics \cite{R5.4, R5.5, R5.6, R5.7}.  

\vspace{0.2cm}
\setlength{\parindent}{0pt}
In non-relativistic quantum mechanics, bound states, scattering states, and resonance states are typically described by time-independent equations in terms of wave functions that depend on the coordinates of individual particles. These equations take into account the free Hamiltonians of the particles and their interaction potentials. In contrast, relativistic quantum mechanics presents considerable challenges in modeling many-body systems \cite{30,31,32}. One significant obstacle is the "many-time problem" \cite{30,31,32}. The many-time problem in relativistic systems arises due to the finite propagation speed of interaction-mediating field carriers (e.g., photons, gluons), which induces retardation effects. Consequently, a single global time parameter cannot be assigned to all interacting particles. To ensure Lorentz covariance and causality, many-body formulations introduce independent time coordinates. Furthermore, managing the total spin of composite systems consisting of multiple spinning particles is a substantial issue. Therefore, formulating equations in this framework requires fully covariant one-time many-body equations that address retardation effects and carefully handle spin algebra \cite{32}. These equations must also account for the general electromagnetic potentials and spinor fields that depend on the spacetime coordinates of the particles. A fully covariant, non-perturbative, and one-time many-body Dirac equation for fermionic systems has been rigorously derived from quantum electrodynamics using the action principle, incorporating both retarded and advanced Green’s functions \cite{32}. Furthermore, this equation can also be obtained as an excited state of Zitterbewegung \cite{33}. Despite this development, finding an exact solution to this equation in $3+1$ dimensions remains difficult, even for well-studied two-body systems like one-electron atoms \cite{34}, positronium, true muonium-like unstable systems, and other $f\overline{f}$ pairs. The primary difficulty lies in the spin algebra, expressed via the Kronecker product of Dirac matrices, leading to a system of 16 radial equations that describe the relative motion of fermion-fermion or $f\overline{f}$ systems after separating radial and angular components. These coupled second-order equations are common in $3+1$ dimensions, and solving them often involves perturbative techniques \cite{34}. However, exact solutions for the fully covariant two-body Dirac equation have been discovered in lower-dimensional systems or for specific $f\overline{f}$ systems with dynamical symmetry \cite{35,36,R5.4,38,39,R5.5,41}. These solutions have been applied to study excitons in monolayer materials \cite{42}, estimate the mass spectra of neutral mesons \cite{35,38}, and explore the thermodynamic properties of interacting $f\overline{f}$ pairs \cite{43}. Moreover, the fully covariant nature of the two-body equation makes it highly effective for studying the evolution of fermion-fermion or $f\overline{f}$ systems in curved spacetime \cite{35,36,39,41,43}. The dynamics of \( f\overline{f} \) pairs within the Dirac framework can be explored using various interaction potentials. The electromagnetic minimal potential couples the Dirac field to an external field through \( \mathcal{A}_\mu = (\phi, \mathbf{A}) \), maintaining gauge and Lorentz invariance. In contrast, the Lorentz scalar potential modifies the mass term \( m \to m + \mathcal{S}(\mathbf{x}) \), altering the particle's effective mass while preserving gauge symmetry. The minimal potential affects the kinetic term, whereas the scalar potential directly alters the intrinsic energy \cite{45,46}.

\vspace{0.2cm}
\setlength{\parindent}{0pt}  
In this paper, we examine coupled $f\overline{f}$ pairs within the framework of position-dependent mass, as formulated in \cite{44,45,46}, by employing the TWH model \cite{AS,47}. Our approach involves solving the corresponding version of the two-body Dirac equation. The manuscript is organized as follows: Section \ref{sec:2} presents the two-body Dirac equation and derives a system of coupled equations that describe the dynamics of $f\overline{f}$ pairs in the TWH framework. In Section \ref{sec:3}, we obtain analytical solutions for these systems under well-known interaction potentials, including Coulomb-type and exponentially decaying Lorentz scalar potentials. Finally, Section \ref{sec:4} summarizes our findings and provides an in-depth discussion of the results.

\section{\mdseries{Wave equation}} \label{sec:2}

\begin{figure}[ht]
\centering
\includegraphics[scale=0.50]{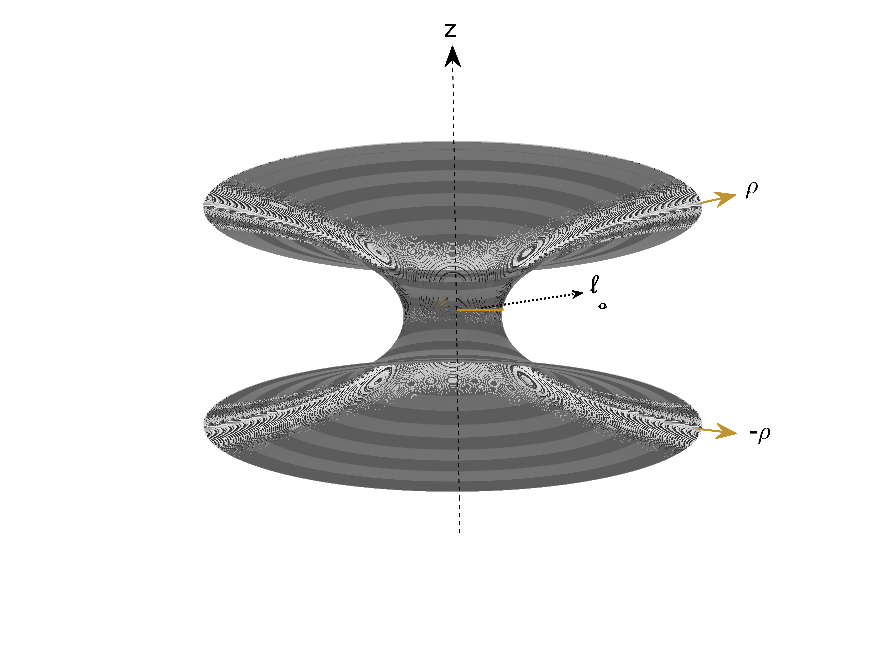}\quad 
\caption{\footnotesize 3D visualization of a wormhole structure described by the parametric expression:  
\(\vec{r}(\rho, \phi) = \sqrt{\rho^2 + \ell_{\circ}^2}(\cos(\phi)\hat{x} + \sin(\phi)\hat{y}) + \ell_{\circ} \sinh^{-1}(\rho/\ell_{\circ})\hat{z}\),  
with \(\ell_{\circ} = 1\) \cite{new-one}. The meridional coordinate \(\rho \in [-5, 5]\) seamlessly connects the lower and upper asymptotically flat regions, demonstrating the geometry's nontrivial topology.}\label{TWH}
\end{figure}

The generalized form of the two-body Dirac equation for a $f\overline{f}$ system with a position-dependent mass is presented in this section. This formulation is derived in the context of a curved spacetime with \(2+1\) dimensions \cite{48}
\begin{equation}
\begin{split}
&\left\lbrace \mathcal{H}^{+} \otimes \gamma^{t^{-}}+\gamma^{t^{+}}\otimes \mathcal{H}^{-} \right\rbrace \Xi\left(x_{+},x_{-} \right)=0,\\
&\mathcal{H}^{\pm}=\left[\gamma^{\mu^{\pm}}\slashed{\nabla}_{\mu_{\pm}} +i\tilde{m}\left(x_{\pm} \right)\textbf{I}_{2}\right],\quad \tilde{m}=\frac{mc}{\hbar},\\
&\slashed{\nabla}_{\mu_{\pm}}=\partial_{\mu_{\pm}}-\Gamma_{\mu_{\pm}}. \label{eq1}
\end{split}
\end{equation}
This equation involves the direct (Kronecker) product, symbolized as $\otimes$, where the subscripts and superscripts $+$ ($-$) are used to distinguish between fermions and antifermions, respectively. The Greek index $\mu$ represents the coordinates within the context of curved spacetime. The constants \(\hbar\) and \(c\) are the Planck constant and the speed of light, as traditionally defined. The rest mass of the particles is denoted by \(m\), while $\textbf{I}_{2}$ refers to the $2 \times 2$ identity matrix. The symbol $\Gamma_{\mu}$ represents the spinorial affine connection associated with the Dirac field, and $\Xi$ is a bi-local spinor that is a function of the spacetime coordinates $\left(x_{+}, x_{-}\right)$ corresponding to the fermions and antifermions. By reintroducing \(c\), the metric that defines the TWH can be rewritten in terms of the line element using the $(+, -, -)$ signature \cite{wh2, AS, yesiltas}:
\begin{eqnarray}
&ds^2 = c^2 dt^2 - d\rho^2 - \left(\rho^2 + \ell_{\circ}^2 \right) d\phi^2, \nonumber \\
&-\infty < t < \infty, \quad \rho \in (-\infty, \infty), \quad \phi \in [0, 2\pi), \label{eq2}
\end{eqnarray}
with the corresponding geometric structure illustrated in Figure \ref{TWH}. 

\begin{figure}[ht]
\centering
\includegraphics[scale=0.40]{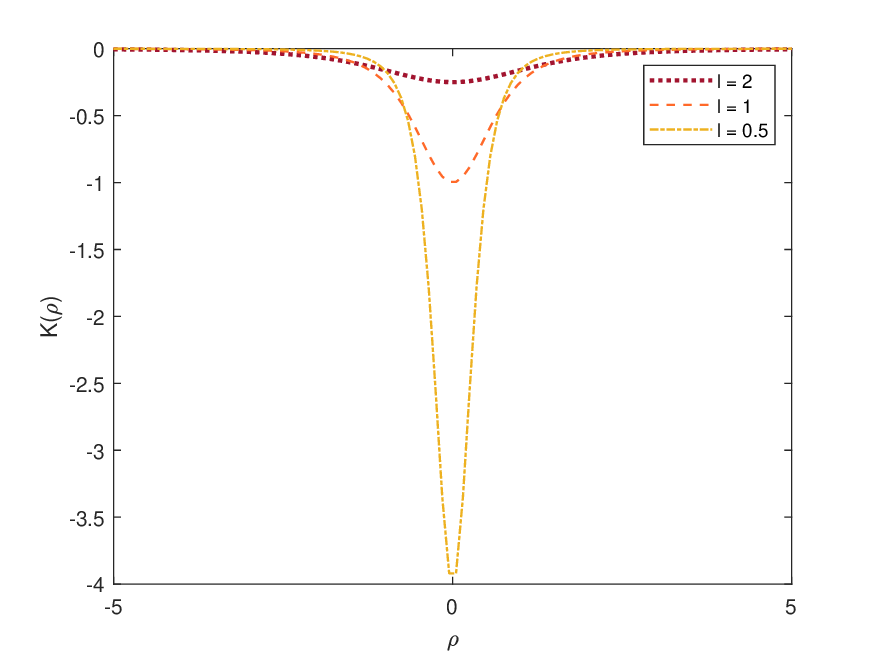}\quad 
\caption{\footnotesize Gaussian curvature \( \mathcal{K} = -\frac{\ell_{\circ}^2}{(\rho^2 + \ell_{\circ}^2)^2} \) plotted as a function of \( \rho \) for \( \ell_{\circ} = 2, 1, 0.5 \). The curvature decreases rapidly as \( \rho \) moves away from zero, with the rate of decrease depending on the value of \( \ell_{\circ} \). Larger values of \( \ell_{\circ} \) result in a flatter Gaussian curvature near the origin, while smaller \( \ell_{\circ} \) values lead to sharper peaks.}\label{GC}
\end{figure}

For this spacetime, the Gaussian curvature ($\mathcal{K}=-\frac{\ell^2_{\circ}}{\rho^2+\ell^2_{\circ}}$) is negative (see Figure \ref{GC}), while the mean curvature is zero \cite{wh2,AS,yesiltas}. 
Using Eq. (\ref{eq2}), the covariant metric tensor is expressed as \( g_{\mu\nu} = \textrm{diag}(c^2, -1, -(\rho^2 + \ell_{\circ}^2)) \). Its inverse is given by \( g^{\mu\nu} = \textrm{diag}(c^{-2}, -1, -(\rho^2 + \ell_{\circ}^2)^{-1}) \). From this metric, the tetrad fields \( e^{a}_{\mu} \) are determined as \( e^{a}_{\mu} = \textrm{diag}(c, 1, \sqrt{\rho^2 + \ell_{\circ}^2}) \) \cite{AS}, and their inverses are \( e_{a}^{\mu} = \textrm{diag}(c^{-1}, 1, (\sqrt{\rho^2 + \ell_{\circ}^2})^{-1}) \). These tetrads satisfy the orthogonality and orthonormality conditions \cite{AS}. Here, Greek indices denote coordinates in curved spacetime (\( \mu = t, \rho, \phi \)), while Latin indices correspond to coordinates in flat Minkowski spacetime (\( a = 0, 1, 2 \)). Given the metric's signature, the flat Dirac matrices \( \gamma^{a} \) are represented in terms of the Pauli matrices (\( \sigma^{x}, \sigma^{y}, \sigma^{z} \)) as follows: \( \gamma^{0} = \sigma^{z} \), \( \gamma^{1} = i\sigma^{x} \), and \( \gamma^{2} = i\sigma^{y} \) \cite{AS}. Consequently, the generalized Dirac matrices for each particle are given by \( \gamma^{t^{\pm}} = \frac{1}{c} \sigma^{z} \), \( \gamma^{\rho^{\pm}} = i\sigma^{x} \), and \( \gamma^{\phi^{\pm}} = i \left(\sqrt{\rho_{\pm}^2 + \ell_{\circ}^2}\right)^{-1} \sigma^{y} \), where \( \gamma^{\mu} = e^{\mu}_{a} \gamma^{a} \). Using the spinorial affine connection relations from \cite{AS}, we find \( \Gamma_{t} = 0 \), \( \Gamma_{\rho} = 0 \), and \( \Gamma_{\phi^{\pm}} = i \frac{\rho_{\pm}}{2 \sqrt{\rho_{\pm}^2 + \ell_{\circ}^2}} \sigma^{z} \). This leads to the result \( \gamma^{\mu^{\pm}} \Gamma_{\mu^{\pm}} = -i \frac{\rho_{\pm}}{2 (\rho_{\pm}^2 + \ell_{\circ}^2)} \sigma^{x} \) \cite{AS}. 

\vspace{0.2cm}

\setlength{\parindent}{0pt} 
Following the standard procedure for two-body problems, we introduce the center of mass motion coordinates (\( \mathcal{R} \)) and relative motion coordinates (\( r \)) for two fermions of equal mass \cite{48}
\begin{eqnarray}
&r_{\mu}=x_{\mu}^{+}-x_{\mu}^{-},\quad \mathcal{R}_{\mu}=\frac{x_{\mu}^{+}}{2}+\frac{x_{\mu}^{-}}{2}, \quad x_{\mu}^{+}=\mathcal{R}_{\mu}+\frac{r_{\mu}}{2},\nonumber\\
& x_{\mu}^{-}=\mathcal{R}_{\mu}-\frac{r_{\mu}}{2},\quad \partial_{x_{\mu}}^{+}=\frac{\partial_{\mathcal{R}_{\mu}}}{2}+\partial_{r_{\mu}},\nonumber\\
&\partial_{x_{\mu}}^{-}=\frac{\partial_{\mathcal{R}_{\mu}}}{2}-\partial_{r_{\mu}},\quad \partial_{x_{\mu}}^{+}+\partial_{x_{\mu}}^{-}=\partial_{\mathcal{R}_{\mu}}. \label{eq3}
\end{eqnarray}
In this study, we explore the relativistic dynamics of coupled $f\overline{f}$ pairs, assuming time-independent interaction potentials. Under this assumption, the spinor field can be decomposed according to Eq. (\ref{eq2}):
\begin{eqnarray*}
\Xi\left(x_{+},x_{-}\right) = \exp\left(-i \frac{\mathcal{E}}{\hbar} t \right) \exp\left(i s \phi \right) 
\begin{pmatrix}
\psi_1(\mathcal{\textbf{R}}, \textbf{r}) \\
\psi_2(\mathcal{\textbf{R}}, \textbf{r}) \\
\psi_3(\mathcal{\textbf{R}}, \textbf{r}) \\
\psi_4(\mathcal{\textbf{R}}, \textbf{r})
\end{pmatrix},
\end{eqnarray*}
where \(\mathcal{E}\) denotes the relativistic energy, and \(s\) represents the total spin of the composite system consisting of a coupled $f\overline{f}$ pair. Upon organizing the system, the resulting set of equations, including an algebraic equation, describes the relative motion of this spinless composite system. The system is characterized by a position-dependent mass \cite{45,46} and is situated within the framework of a TWH, where the center of mass remains at rest at the spatial origin, as detailed in \cite{45,46}
\begin{equation}
\begin{split}
&\tilde{\mathcal{E}}\psi_{+}\left(\rho \right) -\tilde{\mu}\left(\rho\right)\psi_{-}\left(\rho\right)+4\,\hat{\lambda}\,\psi_{2}\left(\rho \right)=0,\\
&\tilde{\mathcal{E}}\psi_{-}\left(\rho \right)-\tilde{\mu}\left(\rho\right)\psi_{+}\left(\rho \right)+8\,\tilde{\eta}\left(\rho\right)\,\psi_{2}\left(\rho \right)=0,\\
&\tilde{\mathcal{E}}\psi_{2}\left(\rho \right)-\hat{\lambda}\,\psi_{+}\left(\rho \right)+2\,\tilde{\eta}\left(\rho\right)\,\psi_{-}\left(\rho \right)=0. \label{eq4}
\end{split}
\end{equation}
The new components are introduced as \(\psi_{\pm} = \psi_1 \pm \psi_4\) and \(\psi_3 = -\psi_2\). In this system of equations, the following substitutions are made:
\begin{equation*}
\begin{split}
&\tilde{\mathcal{E}}=\frac{\mathcal{E}}{\hbar c},\quad \tilde{\mu}\left(\rho\right)=\frac{2mc}{\hbar}+\mathcal{S}\left(\rho\right),\\
&\hat{\lambda}=\partial_{\rho}+\frac{\rho}{\left(4\ell_{\circ}^2+\rho^2 \right)},\quad \tilde{\eta}\left(\rho\right)=\frac{s}{\sqrt{\rho^2+4\ell^{2}_{\circ}}}.
\end{split}
\end{equation*}
Equation (\ref{eq4}) shows that for \(s = 0\), all spinor components can be expressed in terms of \(\psi_{+}\) (see \cite{AS} for \(s \neq 0\) with \(\tilde{\mu}(\rho) = 0\)). Therefore, for a spinless composite system (\(s = 0\)), the following non-perturbative wave equation is obtained:
\begin{equation}
\begin{split}
&\partial^{2}_{\rho}\psi_{+}+\frac{2\rho}{\rho^2+4\ell_{\circ}^2}\partial_{\rho}\psi_{+}+\left[\frac{\tilde{\mathcal{E}}^2-\tilde{\mu}\left(\rho\right)^{2}}{4}+\mathcal{C}\left(\rho\right)\right]\psi_{+}=0, \label{WE}\\
&\mathcal{C}\left(\rho\right)=\frac{1}{\rho^2+4\ell_{\circ}^2}-\frac{\rho^2}{(\rho^2+4\ell_{\circ}^2)^2}.
\end{split}
\end{equation}
The wave equation can be recast into a textbook one-dimensional radial Schrödinger-like form by employing the ansatz \(\psi_{+}(\rho)=\frac{1}{\sqrt{\rho^2+4\ell_{\circ}^2}}\,\psi(\rho)\), to yield the following radial equation:
\begin{equation}
\begin{split}
&\partial^{2}_{\rho}\psi\left(\rho\right)+\left[\epsilon-V_{\text{eff}}\left(\rho\right)\right]\psi\left(\rho\right)=0,\\
&\epsilon=\tilde{\mathcal{E}}^2/4,\quad V_{\text{eff}}\left(\rho\right)=\tilde{\mu}\left(\rho\right)^2/4.\label{SCH-form}
\end{split}
\end{equation}
This formulation indicates that while the wave function \(\psi_{+}(\rho)\) is influenced by the radius of the wormhole throat, the energy spectrum of the system remains unaffected by this variable. Consequently, the energy properties of the system are governed primarily by the Lorentz scalar potential \(\mathcal{S}(\rho)\) and the rest mass energy.

\section{\mdseries{Exact solutions}} \label{sec:3}

In this section, we seek to derive analytical solutions for the equation (\ref{SCH-form}) by considering two distinct forms of the Lorentz scalar potential \(\mathcal{S}(\rho)\).

\subsection{\mdseries{Exact solutions for $\mathcal{S}(\rho) = -\frac{\alpha}{\rho}$}} \label{sec:3:1}
\begin{figure*}[ht]
    \centering
    \subfigure[$\ell_{\circ}=1$]{\includegraphics[scale=0.35]{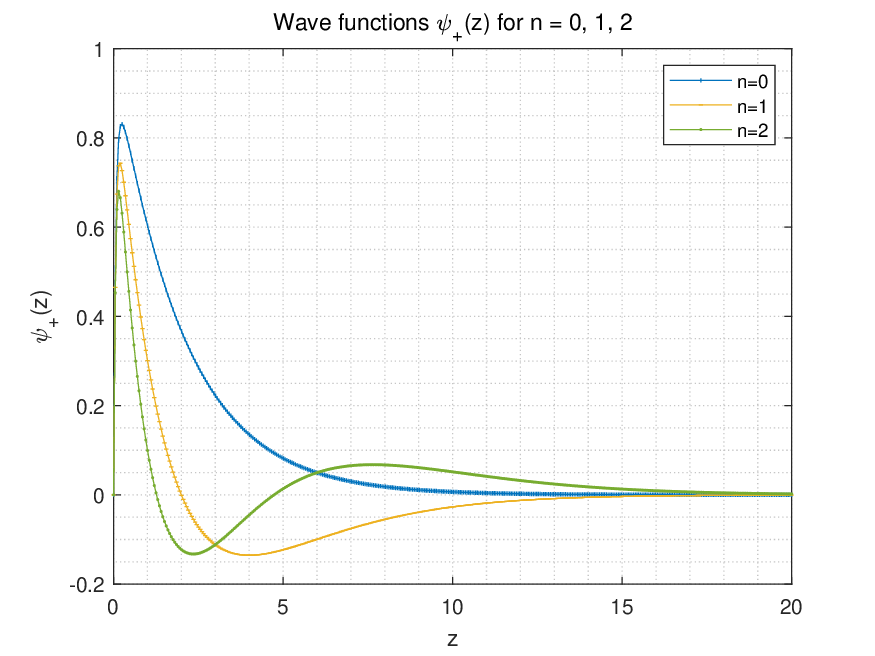}}\quad 
    \subfigure[$\ell_{\circ}=10$]{\includegraphics[scale=0.35]{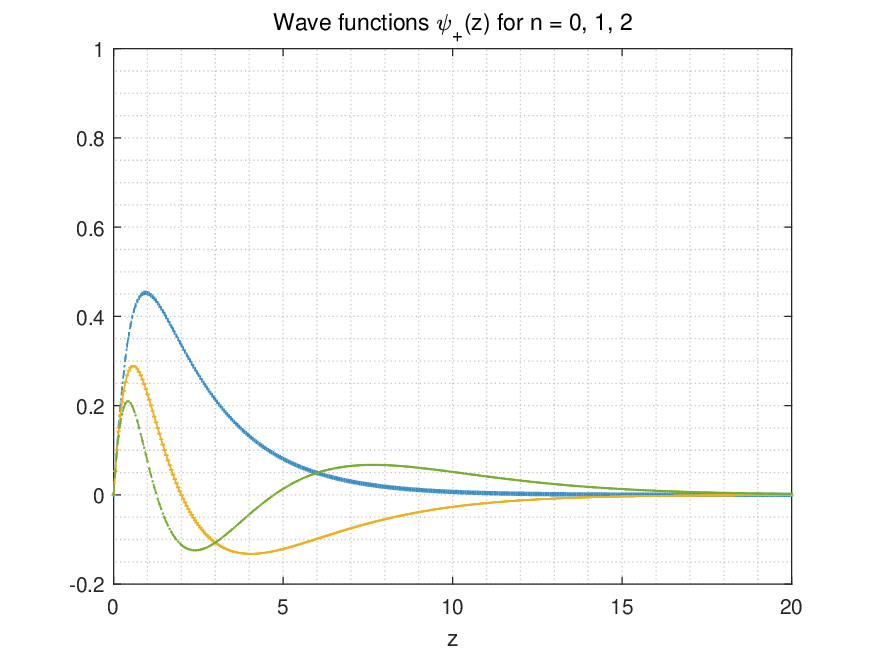}}\\
    \subfigure[$\ell_{\circ}=1$]{\includegraphics[scale=0.35]{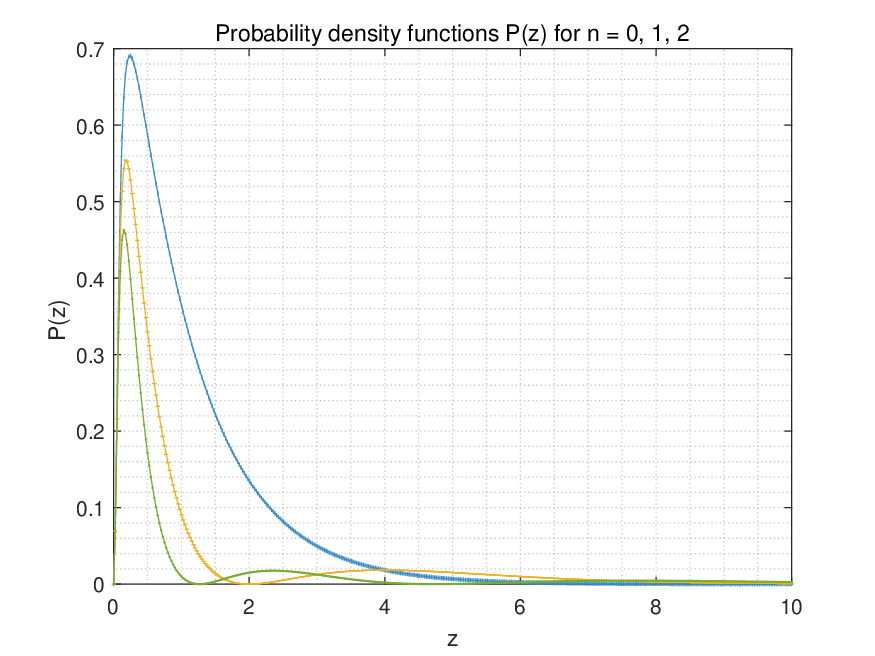}}\quad 
    \subfigure[$\ell_{\circ}=10$]{\includegraphics[scale=0.35]{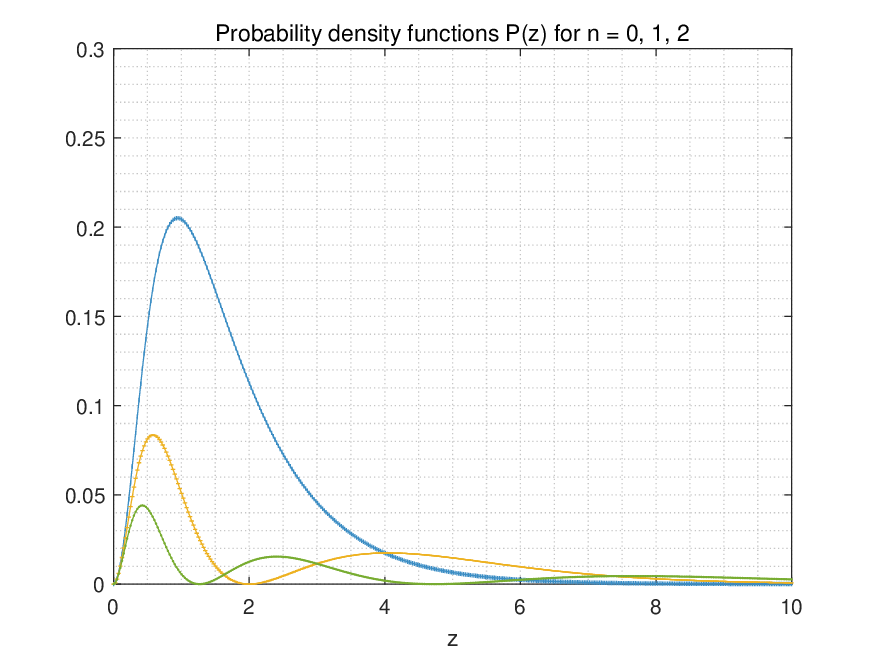}} 
    \caption{\footnotesize Wave functions \(\psi_{+}(z)\) and probability density functions \(P(z)\) for quantum states \(n = 0\), \(n = 1\), and \(n = 2\), computed using the parameters \(N = 1\), \(M = 1\), and \(\tilde{\mathcal{E}} = 0.999\). The parameter \(\tilde{b}\) is given by \(\tilde{b} = 1 + \sqrt{1 + \alpha^2}\), where \(\alpha\) denotes the fine-structure constant \(\alpha = 1/137\). The plots illustrate \(\psi_{+}(z)\) and \(P(z)\) as functions of \(z\), highlighting the distinct features of the wave functions and probability densities for several \(\ell_\circ\) values.}
    \label{figa}
\end{figure*}
When the Lorentz scalar potential is specified as above, the wave equation takes the following form of the one-dimensional Coulombic radial Schr\"{o}dinger equation:  
\begin{equation}
\partial^{2}_{\rho} \psi(\rho) + \left(\frac{\tilde{\mathcal{E}}^2-M^2}{4} + \frac{M\alpha}{2\rho} - \frac{\alpha^2}{\rho^2} \right) \psi(\rho) = 0,
\end{equation}
where \(M = \frac{2mc}{\hbar}\). By applying the substitution \(z = \rho \sqrt{M^2 - \tilde{\mathcal{E}}^2}\), the equation is transformed into the following form:
\begin{equation}
\begin{split}
&\partial^{2}_{z} \psi(z) + \left(-\frac{1}{4} + \frac{q}{z} + \frac{\frac{1}{4} - p^2}{z^2} \right) \psi(z) = 0,\\
&q = \frac{M\alpha}{2\sqrt{M^2-\tilde{\mathcal{E}}^2}},\quad p = \frac{\sqrt{1+\alpha^2}}{2}.
\end{split}
\end{equation}
This equation is recognized as the well-known Whittaker differential equation \cite{36}. Its regular solution near the singular point \(z = 0\) can be expressed in terms of the confluent hypergeometric function \({}_1F_1\), as follows \cite{36}:
\begin{equation}
\begin{split}
&\psi(z) = N\,e^{-\frac{z}{2}} z^{\frac{\tilde{b}}{2}} \, {}_1F_1\left(\tilde{a}; \tilde{b}; z\right),\\
&\tilde{a} = \frac{1}{2} + p - q = \frac{1+\sqrt{1+\alpha^2}}{2} - \frac{M\alpha}{2\sqrt{M^2-\tilde{\mathcal{E}}^2}},\\
&\tilde{b} = 1 + 2p = 1 + \sqrt{1+\alpha^2}.\label{WF1}
\end{split}
\end{equation}
It is obvious that the radial wave function above converges to zero as \(z = 0 = \rho\) (a condition that is manifestly introduced by the central repulsive core \(\sim 1/z^2\)) and as \(z = \infty = \rho\). This ensures that the system forms bound states and the wave functions remain finite. Moreover, the truncation of the confluent hypergeometric series to a polynomial of degree \(n \geq 0\) \cite{36} enforces the quantization condition \(\tilde{a} = -n\) \cite{36} and allows the square-integrability of the radial wave function. As a result, the energy spectrum (\(\mathcal{E} \rightarrow \mathcal{E}_n\)) is given by:
\begin{equation*}
\mathcal{E}_{n} = \pm 2m c^2 \sqrt{1 - \frac{\alpha^2}{(2n + \sqrt{1+\alpha^2} + 1)^2}}.  
\end{equation*}
This energy expression can be expanded as a power series, characteristic of single-electron systems:
\begin{equation}
\mathcal{E}_{n} \approx 2mc^2 \left\lbrace 1 - \frac{\alpha^2}{8\,(n+1)^2} + \frac{(8n+7)\,\alpha^4}{128\,(n+1)^4} - \mathcal{O}(\alpha^6) \right\rbrace .\label{spec-1}
\end{equation}
In the relativistic energy spectrum, the leading term represents the total rest energy (\(2mc^2\)), while the \(\alpha^2\)-dependent term corresponds to the non-relativistic binding energy. For para-positronium, the singlet spin state of the electron-positron pair, the binding energy in the ground state is given by \(\mathcal{E}^b_0 \approx -m_e c^2 \alpha^2/4 \sim -6.803\) eV\(\footnote{Here, we use \(m_e = 9.10938356 \times 10^{-31}~\text{kg}\), \(c = 2.99792458 \times 10^{8}~\text{m/s}\), and \(\alpha = 1/137.035999\), the fine-structure constant.}\). The higher-order terms (\(\alpha^4, \alpha^6\)) account for relativistic corrections.

Returning to the associated wave equation, we can relate \(\psi(z)\) to the wave function \(\psi_{+}(z)\) as follows:
\begin{equation}
\begin{split}
&\psi_{+}(z) = N\,\frac{z^{\frac{\tilde{b}}{2}}}{\sqrt{4\ell_\circ^2\left(M^2-\tilde{\mathcal{E}}^2\right) + z^2}}\,e^{-\frac{z}{2}} \, {}_1F_1\left(\tilde{a}; \tilde{b}; z\right).\label{WF2}
\end{split}
\end{equation}
This expression reveals that the wave function is sensitive to variations in the wormhole throat radius \(\ell_\circ\), even though the energy profile of the bound pair remains unaffected by it. The spatial configuration and probability distribution of the pair are influenced by the wormhole geometry due to this dependence. Since the total rest energy (\(2mc^2\)) is always greater than the relativistic energy (\(\mathcal{E}_n\)) and the binding energy \(\mathcal{E}^b_n(\alpha)\) is negative for all quantum states, the probability density \(P(z) = \int_{-\infty}^\infty |\psi_{+}(z)|^2 \, dz\):
\begin{equation}
P(z) = |N|^2 \int_{-\infty}^\infty \frac{z^{\tilde{b}}}{4\ell_\circ^2\left(M^2-\tilde{\mathcal{E}}^2\right) + z^2} e^{-z} \left|{}_1F_1\left(\tilde{a}; \tilde{b}; z\right)\right|^2 dz,\label{PDF-1}
\end{equation}
depends inherently on the wormhole throat radius. The dependence of the wave functions and the probability density functions is illustrated in Figure \ref{figa}.

\subsection{\mdseries{Exact solutions for $\mathcal{S}(\rho)=-A\,\exp(-\rho/\lambda_{c})$}} \label{sec:3:3}

In this scenario, the wave equation can be written as  
\begin{equation}  
\frac{d^2 \psi(\rho)}{d \rho^2} + \left( \frac{\tilde{\mathcal{E}}^2}{4} - \frac{\left(M - A\, e^{-\frac{\rho}{\lambda_c}} \right)^2}{4} \right) \psi(\rho) = 0,  
\end{equation}  
where \( \lambda_c \) denotes the Compton wavelength associated with fermions \cite{book1,book2,49}. By making the substitution \( x = A \lambda_c e^{-\rho / \lambda_c} \), the equation is reformulated and, using the ansatz \(\psi(x) = \frac{\tilde{\psi}(x)}{\sqrt{x}}\), it reduces to the Whittaker wave equation. The resulting expression takes the form  
\begin{equation}  
\begin{split}  
&\partial_x^2 \tilde{\psi}(x) + \left(-\frac{1}{4} + \frac{q_1}{x} + \frac{\frac{1}{4} - p_1^2}{x^2} \right) \tilde{\psi}(x) = 0, \\  
&q_1 = \frac{\lambda_c}{2}M, \quad p_1 = \frac{\lambda_c}{2} \sqrt{M^2 - \tilde{\mathcal{E}}^2}.  
\end{split}  
\end{equation}  
By following the procedure outlined in Section \ref{sec:3:1}, the regular solution to this equation can be expressed in terms of the confluent hypergeometric function of the first kind (refer to \cite{49} for further details):  
\begin{equation}  
\begin{split}  
&\tilde{\psi}(x) = N_1 e^{-\frac{x}{2}} x^{\frac{\tilde{b}_1}{2}} \, {}_1F_1\left(\tilde{a}_1; \tilde{b}_1; x\right), \\  
&\tilde{a}_1 = \frac{1}{2} - \frac{\lambda_c}{2}\sqrt{M^2 - \tilde{\mathcal{E}}^2} - \frac{\lambda_c}{2}M, \\  
&\tilde{b}_1 = 1 + \lambda_c \sqrt{M^2 - \tilde{\mathcal{E}}^2}.  
\end{split}  
\end{equation}  
From this solution, the energy spectrum can be derived as  
\begin{equation}  
\mathcal{E}_n = \pm i\frac{\hbar c}{\lambda_c} \sqrt{(2n + 1)^2 - 2M\lambda_c(2n + 1)}.  
\end{equation}  
This exact result reveals that the system exhibits decay behavior over time, with the corresponding decay time expressed as (see also \cite{36,49})  
\begin{equation}  
\tau_n = \frac{\hbar}{|\Im \mathcal{E}_n|} = \frac{\lambda_c / c}{\sqrt{(2n + 1)^2 - 2M\lambda_c(2n + 1)}},  
\end{equation}  
provided that \( n + \frac{1}{2} > M\lambda_c \), as the wave function exhibits time dependence \( \Xi \propto e^{-i\frac{\mathcal{E}_n}{\hbar} t} \). This further suggests that the decay rate is influenced by both the Compton wavelength and the total mass of the particles \footnote{Note that \( M = \frac{2mc}{\hbar} \), and \( M\lambda_c \) is dimensionless since \( \lambda_c = \frac{\hbar}{mc} \).}. In the case where \( n + \frac{1}{2} < M\lambda_c \), the energy takes the form  
\begin{equation}  
\mathcal{E}_n = \pm \frac{\hbar c}{\lambda_c} \sqrt{2M\lambda_c(2n + 1) - (2n + 1)^2},  
\end{equation}  
indicating oscillatory behavior of the system without energy dissipation. A particular case arises when the fermions attain a critical mass, expressed as  
\begin{equation}  
m_c(n) = \frac{\tilde{n} \hbar}{2 \lambda_c c}, \quad \tilde{n} = n + \frac{1}{2},  
\end{equation}  
at which point the energy vanishes, suggesting that the system ceases to evolve over time \cite{50}, although it remains present in space. Returning to the wave function \(\psi_+(x)\), the solution takes the form  
\begin{equation}  
\psi_+(x) = N_1 \frac{x^{\frac{\tilde{b}_1}{2}}}{\sqrt{\ln\left(\frac{x}{A\lambda_c}\right)^2 \lambda_c^2 x + 4\ell_\circ^2 x}} e^{-\frac{x}{2}} \, {}_1F_1\left(\tilde{a}_1; \tilde{b}_1; x\right).\label{WF-3}  
\end{equation}  
As \( x \to 0 \), the term \( x^{\frac{\tilde{b}_1}{2}} \) behaves smoothly due to \( \tilde{b}_1 > 0 \), while the denominator \( \sqrt{\ln\left(\frac{x}{A \lambda_{c}}\right)^2 \lambda_{c}^2 x + 4 \ell_{\circ}^2 x} \) is dominated by the term \( 4 \ell_{\circ}^2 x \), leading to \( 2 \ell_{\circ} \sqrt{x} \). The exponential term \( e^{-\frac{x}{2}} \) tends to 1, and the confluent hypergeometric function \( {}_1F_1(\tilde{a}_1; \tilde{b}_1; x) \) reduces to 1, as \( {}_1F_1(\tilde{a}_1; \tilde{b}_1; 0) = 1 \). By combining these factors, we obtain \( \psi_{+}(x) \sim \frac{x^{\frac{\tilde{b}_1-1}{2}}}{2 \ell_{\circ}} \), which implies that \( \psi_{+}(x) \to 0 \) as \( \tilde{b}_1 > 1 \).

As \( x \to \infty \), the term \( x^{\frac{\tilde{b}_1}{2}} \) increases and the denominator is dominated by \( \ln\left(\frac{x}{A \lambda_{c}}\right)^2 \lambda_{c}^2 x \). The exponential factor \( e^{-\frac{x}{2}} \) decays exponentially. The confluent hypergeometric function \( {}_1F_1(\tilde{a}_1; \tilde{b}_1; x) \) behaves asymptotically as \( \frac{\Gamma(\tilde{b}_1)}{\Gamma(\tilde{a}_1)} e^x x^{\tilde{a}_1 - \tilde{b}_1} + \frac{\Gamma(\tilde{b}_1)}{\Gamma(\tilde{b}_1 - \tilde{a}_1)} x^{-\tilde{a}_1} \), where the first term dominates for large \( x \). However, the exponential growth of \( {}_1F_1 \) is suppressed by the factor \( e^{-\frac{x}{2}} \), leading to \( \psi_{+}(x) \sim e^{-\frac{x}{2}} x^{\tilde{a}_1 - \frac{\tilde{b}_1}{2}} \). This ensures that \( \psi_{+}(x) \) decays exponentially to zero as \( x \to \infty \), noting that \( \frac{2\tilde{a}_1 - \tilde{b}_1}{2} < 0 \).

It is also worth emphasizing that the probability function becomes time-dependent, in addition to its spatial dependence, if \( n + \frac{1}{2} > 2 \), since \( \Xi \propto e^{-i \frac{\mathcal{E}_{n}}{\hbar} t} \). These results further indicate that the probability density function is sensitive to variations in the wormhole throat radius.

\section{\mdseries{Summary and discussions}} \label{sec:4}

This study investigates the dynamics of $f\overline{f}$ pairs within a TWH spacetime by solving the covariant two-body Dirac equation, which incorporates a position-dependent mass, \(m \to m(r)\). Focusing on a static, radially symmetric (2+1)-dimensional TWH characterized by a constant redshift function and a specific shape function, we examine two distinct Lorentz scalar potentials: (i) a Coulomb-like potential and (ii) an exponentially decaying potential.

In the first case (i), we derive exact solutions, leading to the following energy spectrum:
\[
\mathcal{E}_n \approx 2mc^2 \left[ 1 - \frac{\alpha^2}{8\,(n+1)^2} + \frac{(8n+7)\,\alpha^4}{128\,(n+1)^4} - \mathcal{O}(\alpha^6) \right],
\]
where the leading term, \(2mc^2\), represents the total rest energy, while the term proportional to \(\alpha^2\) corresponds to the non-relativistic binding energy. For para-positronium in its singlet spin state, the ground-state binding energy is computed as \(\mathcal{E}_0^b \approx -m_e c^2 \alpha^2/4 \sim -6.803\) eV, with higher-order corrections (\(\alpha^4, \alpha^6\)) providing relativistic adjustments. The same energy spectrum applies to excitonic states in analogous condensed matter systems when the substitution \(\alpha \to \alpha/\epsilon_{\text{eff}}\) is made \cite{exciton}, where \(\epsilon_{\text{eff}}\) denotes the effective dielectric constant of the surrounding medium. Under this assumption, the binding energy values for the excitonic states are presented in Table \ref{table:1}.
\begin{table}[h!]
\centering
\scalebox{0.7}{
\begin{tabular}{|c|c|c|c|c|}
\hline
$\epsilon_{eff}$ & $\mathcal{E}^b_0$ (eV) & $\mathcal{E}^b_1$ (eV) & $\mathcal{E}^b_2$ (eV) & $\mathcal{E}^b_3$ (eV) \\ \hline
2.5 & -1.0885 & -0.2721 & -0.1209 & -0.0680 \\ \hline
3 & -0.7559 \cite{e-new-1} & -0.1890 & -0.0840 & -0.0472 \\ \hline
3.5 & -0.5553 & -0.1388 & -0.0617 & -0.0347 \\ \hline
4 & -0.4252 & -0.1063 & -0.0472 & -0.0266 \\ \hline
4.5 & -0.3359 \cite{e-new-2} & -0.0840 & -0.0373 & -0.0210 \\ \hline
5 & -0.2721 & -0.0680 & -0.0302 & -0.0170 \\ \hline
\end{tabular}}
\caption{\footnotesize Binding energy levels, \(\mathcal{E}^b_n\), of excitonic states for various \(\epsilon_{\text{eff}}\) values.}
\label{table:1}
\end{table}
Notably, the energy spectrum for this coupled $f\overline{f}$ system remains independent of the wormhole parameters, determined by the interaction potential. However, the wave function (see Eq. (\ref{WF2})) depends on the wormhole geometry, indicating that while the wormhole structure does not modify the energy levels, it does influence the spatial localization and dynamics of the quantum system. This spatial dependence has implications for quantities such as spatial correlations and probability densities (see Eq. (\ref{PDF-1})). These results demonstrate that the geometry of the wormhole subtly affects quantum states without altering the corresponding energy eigenvalues. As illustrated in Figure \ref{figa}, the probability density and wave function peaks shift as the wormhole throat radius varies. For smaller throat radii, the peaks are localized closer to the wormhole throat, where the gravitational potential is strongest, promoting confinement near the origin. Conversely, larger throat radii result in peaks shifting outward, reflecting reduced confinement and allowing the quantum states to spread more broadly. This behavior reveals how the wormhole throat radius influences the spatial configuration of quantum states.

In the second scenario (ii), involving an exponentially decaying potential, we derive exact solutions that yield the following energy expression:
\[
\mathcal{E}_n = \pm \frac{\hbar c}{\lambda_c} \sqrt{2M\lambda_c(2n+1) - (2n+1)^2}.
\]
This result identifies critical mass thresholds, \(m_c = \frac{(n+\frac{1}{2})\hbar}{2\lambda_c c}\), at which the system transitions to zero-energy states, resulting in vanishing wave functions over time. Stability is achieved for \(n+\frac{1}{2} < 2\), corresponding to bound states that do not decay, while \(n+\frac{1}{2} > 2\) describes unstable states that decay over time, allowing modes to escape the wormhole. For example, the ground state (\(n=0\)) and the first excited state (\(n=1\)) remain bound, whereas higher states (\(n > 2\)) represent leaking modes with wave functions that vanish as \(x \to \infty\). Similar to the previous scenario, the energy spectrum in this context is determined exclusively by the system's intrinsic characteristics, while the spatial configuration of the wave function is shaped by the wormhole's geometry. 
\vspace{0.2cm}

\setlength{\parindent}{0pt} 
The spatial features affect probability densities and the localization of quantum states, emphasizing the significance of the wormhole's throat in influencing quantum behavior. These results offer important perspectives on many-body quantum dynamics in curved spacetimes and suggest parallels with condensed matter systems, where the use of an effective dielectric constant or substitution \(c \to v_F\) (Fermi velocity) could lead to a deeper understanding of many-body states in analogous geometries. 

\section*{\small{CRediT authorship contribution statement}}

\textbf{Abdullah Guvendi}:
Conceptualization, Methodology, Investigation, Writing – Review and Editing, Formal Analysis, Validation, Visualization, Project Administration.

\textbf{Omar Mustafa}: Conceptualization, Methodology, Investigation, Writing – Review and Editing, Formal Analysis, Validation, Visualization.

\textbf{Semra Gurtas Dogan}: Conceptualization, Methodology, Investigation, Writing – Review and Editing, Formal Analysis, Validation, Visualization.

\section*{\small{Funding}}
No financial support was received for this research.

\section*{\small{Data availability}}
All relevant data from this study have been integrated into this article.

\section*{\small{Competing Interests }}
The authors state that there are no conflicts of interest.

\end{document}